\documentclass[preprint,showpacs,preprintnumbers,amsmath,amssymb]{revtex4}

\usepackage{graphicx}
\usepackage{dcolumn}
\usepackage{bm}

\def\Kbar{$\overline{\rm K}$}
\def\hv{{\it h}$\nu$}

\begin{document}

\title{Rashba splitting of 100 meV in Au-intercalated graphene on SiC}

\author{D. Marchenko$^{1}$, A. Varykhalov$^{1}$, J. S\'anchez-Barriga$^{1}$, Th. Seyller$^{2}$ and O. Rader$^{1}$}
\affiliation{{ }$^{1}$ Helmholtz-Zentrum Berlin f\"ur Materialien und Energie, 
Elektronenspeicherring BESSY II, Albert-Einstein-Stra\ss e 15,  12489 Berlin, Germany}
\affiliation{{ }$^{2}$ Institut f\"ur Physik, Technische Universit\"at Chemnitz, Reichenhainer Strasse 70, 09126 Chemnitz, Germany}

\begin{abstract}
Intercalation of Au can produce giant Rashba-type spin-orbit splittings in graphene 
but this has not yet been achieved on a semiconductor substrate. 
For graphene/SiC(0001), Au intercalation yields two phases with different doping. 
Here, we report the preparation of an almost pure p-type graphene phase after Au intercalation. 
We observe a 100 meV Rashba-type spin-orbit splitting at 0.9 eV binding energy. 
We show that this giant splitting is due to hybridization and much more limited in energy and momentum space than for Au-intercalated graphene on Ni. 
\end{abstract}


\maketitle

In a future spintronics, graphene could be used in various ways: as efficient leads for spin currents due to low spin-orbit interaction and large spin relaxation lengths \cite{TombrosNature2007}, as ferromagnetic graphene due to a proximity magnetization \cite{DedkovHornAPL10}, as spin filter when a large Rashba type spin-orbit interaction and a band gap at the Dirac point are imposed by a substrate \cite{VarykhalovNatComm15}, or when the spin-orbit interaction is strongly enhanced by covalently bonded adatoms \cite{OzyilmazNatPhys2013}.
The Rashba effect on a graphene Dirac cone leads to a complex band topology with a gapped and an ungapped cone similar to spinless bilayer graphene \cite{RashbaPRB09,GmitraPRB09}.
It has been shown by ARPES that for Au-intercalated graphene/Ni(111) the Dirac cone is intact \cite{VarykhalovPRL08} and gapless \cite{VarykhalovPRB10}. 
A giant Rashba splitting of $\sim100$ meV was found for graphene/Au/Ni(111) \cite{MarchenkoNatComm12}.  
On light elements, the spin-orbit splitting of graphene cannot be resolved in experiment, this is the case with Ni(111) and Co(0001) \cite{RaderPRL09,VarykhalovPRB09}. Where, instead, an unexpected Dirac point forms at higher binding energy as part of a pair of bonding and antibonding Dirac cones \cite{VarykhalovPRX12}. These cones have been found to be highly spin polarized by the exchange interaction from the substrate \cite{MarchenkoPRB15,UsachovNL15}. 
 
The interfaces of graphene with metals are structurally rather well defined, and the intercalation of Au is driven by the transition metal substrate. The creation of the spin-orbit interaction in the graphene is well understood based on spin-dependent hybridization with d-states of the substrate. This is the case for graphene/Au/Ni(111) \cite{MarchenkoNatComm12} and is confirmed indirectly by the vanishing spin-orbit splitting of the graphene-Bi interface because atomic numbers of Bi and Au are similar but hybridization with d-states is possible only for Au \cite{ShikinNJP13}. A similar hybridization causes the $\sim50$ meV spin-orbit splitting for graphene/Ir(111) \cite{MarchenkoPRB13ir}.

To make the aforementioned effects useful for spintronics, they have to be confirmed on an insulating substrate such as SiC because otherwise the metallic substrate will dominate charge and spin transport. On bare SiC, graphene does not show an enhanced spin-orbit splitting as has been clarified by spin-resolved photoemission \cite{MarchenkoPRB13sic}. Recently, a spin-orbit splitting of 17 meV has been measured for graphene on the semiconductor WS$_2$ \cite{OzyilmazNCOMM2014}.

In the context of a possible p-doping of graphene by metals, the intercalation of Au under graphene on SiC has already been studied \cite{PremlalAPl09,Gierz10}, which serves as starting point for our search for enhanced spin-orbit splittings in this system. Scanning tunneling microscopy (STM) and spectroscopy (STS) \cite{PremlalAPl09} reveal two stages of Au intercalation: intercalation of Au clusters without change of the n-type doping of graphene on SiC and intercalation of 1 monolayer (ML) of Au with a $2\sqrt{3} \times 2\sqrt{3} R30^\circ$-Au(111) structure. It was speculated whether the latter phase shows p-doping \cite{PremlalAPl09}. 
The existence of an n-doped phase shifted by $\sim-0.85$ eV and a slightly p-doped phase has been confirmed subsequently by ARPES performed in the near-Fermi-energy region \cite{Gierz10}. 

In the present work, we study Au intercalation by spin-ARPES and core-level photoemission. We confirm the n- and p-doped phases and study how the hybridization between the Dirac cone and Au states changes between the two phases. The p-doped phase shows a metallic Au $sp$-band. We find a giant Rashba-type spin-orbit splitting only in a hybridization region. 

Measurements have been performed at BESSY II with a hemispherical analyzer coupled to a Rice University Mott-type spin detector sensitive to the in-plane spin component. 
Overall energy and angular resolution of the experiments was 80 meV and 1$^\circ$.
Measurements have been conducted at room temperature. 

Like in previous work \cite{MarchenkoPRB13sic}, we start out with a structural graphene monolayer that does not show the graphene band dispersion (so-called zero-layer graphene). It is grown by annealing on the Si-face of nitrogen-doped 6H-SiC(0001) as described in Refs. \onlinecite{MarchenkoPRB13sic} and \onlinecite{Emtsev09}. Au is then deposited on the sample at room temperature followed by annealing to 800$^\circ$C. This leads to n-type graphene and, after repeating the deposition and annealing procedure several times, to the p-type graphene phase. It is particularly difficult to reach a pure p-type phase, and it cannot be excluded that excess Au is produced when the pure p-phase is reached. 

Figure 1 shows Au $4f$ and Si $2p$ core level spectra for the zero-layer graphene and the two differently doped graphene phases. The Au $4f$ core levels compare rather well to previous work \cite{Gierz10} in terms of overall shape and energy shift. It is also seen that the p-type phase is not pure. 

This can also be deduced from the ARPES data in Fig. 2. To facilitate the spin-resolved experiment, the data of Fig. 2 have been taken with the same instrument meaning the same energy and angle resolution. In Fig. 2a we see the n-type graphene/Au/SiC with a shifted and gapped Dirac cone ($\pi$-band bottom at 9.5 eV binding energy). Most prominent hybridizations with Au $5d$ states appear around 5-6 eV. Figure 2b shows the p-type graphene/Au/SiC with additional hybridizations around 3$-$4 eV and changes in the Au band dispersion. In particular, the Au $6sp$ band has changed and shows a metallic behaviour with a pronounced Fermi-level crossing. This is not observed in graphene/Au/Ni(111) because the Au $6sp$-states strongly hybridize with the Ni substrate, as is confirmed by band structure calculations \cite{MarchenkoNatComm12}. 

The zoom of the vicinity of the Dirac point of the p-type sample in Fig. 2d reveals a faint shadow of the shifted n-type Dirac cone. More importantly, very prominent extra hybridizations appear at 1.1 and 1.9 eV binding energy that do not exist for the n-type graphene (Fig. 1c). We assign these hybridizations to the influence of the Au. 

Because of the larger density of Au in the p-type phase and possibly enhanced spin-orbit interaction effects, we study this phase by spin-resolved photoemission. In the case of Au-intercalated graphene on Ni(111), the spin splitting of the $\pi$-band stays constant at $\sim100$ meV from the Dirac point near the Fermi energy to the hybridizations with Au $5d$ states above 4 eV binding energy where the splitting was further enhanced to values of several hundreds meV \cite{MarchenkoNatComm12}. This large energy range of high and constant splitting has been confirmed by calculations \cite{MarchenkoNatComm12}. In the present case, we cannot resolve such a giant Rashba splitting anywhere near the Fermi energy. 

Figure 3 shows spin-resolved measurements for p-type graphene in the region of hybridization with the flat band at 1.1 eV binding energy. The measured region is marked on Fig. 2d by white solid line. Because the spectrum cuts through the hybridization gap (gap size $\sim300$ meV), two bands are probed. The upper band at lower binding energy exhibits a giant spin splitting of 100 meV whereas the lower band appears not split. Similarly to previous work \cite{MarchenkoPRB13sic}, we estimate the lower limit of the detectable splitting to 20 meV. The difference to the measurements of graphene/Au/Ni(111) \cite{MarchenkoNatComm12}, which showed a giant splitting, is large. We can compare to the band structure calculation of graphene in contact with 1 ML Au (see Fig. S11 in the Supplementary Material of Ref. \cite{MarchenkoNatComm12}). This calculation showed a similar Au $6sp$ band as in Fig. 2b. The calculated Rashba splitting for this configuration involving a full Au monolayer was only of the order of 10 meV \cite{MarchenkoNatComm12}, consistent with the present measurements. Interestingly, a tenfold enhancement was predicted for the hybridization point of graphene/SiC \cite{MarchenkoPRB13sic}. If we apply this enhancement not to the low (0.05 meV) Rashba spin-orbit splitting of graphene/SiC but to the interaction with the Au monolayer observed around 1.1 eV binding energy, a value of 100 meV appears reasonable. 

We have shown that the previously observed p-type phase of Au-intercalated graphene is characterized by a metallic Au $6sp$ band and various unexpected hybridizations with flat bands. We find that the Rashba-type spin-orbit splitting reaches giant values of $\sim100$ meV only near this hybridization region which points to a different interface between graphene and Au with weaker hybridization in the present system. Our work shows that there are two strategies to reach 100 meV spin-orbit splitting of graphene on insulating SiC: modification of the growth to move individual Au atoms closer to the graphene as occurs with graphene/Au/Ni(111) \cite{MarchenkoNatComm12} or modification of the charge doping to have regions of strong hybridization at the Fermi energy.

{\it Acknowledgement.} We thank Felix Fromm for providing zero-layer graphene/SiC samples. This work was supported by SPP 1459 of the Deutsche  Forschungsgemeinschaft.

\newpage

%
%
%
%
%
%


\begin{figure}[h] \includegraphics[width=0.9\textwidth]{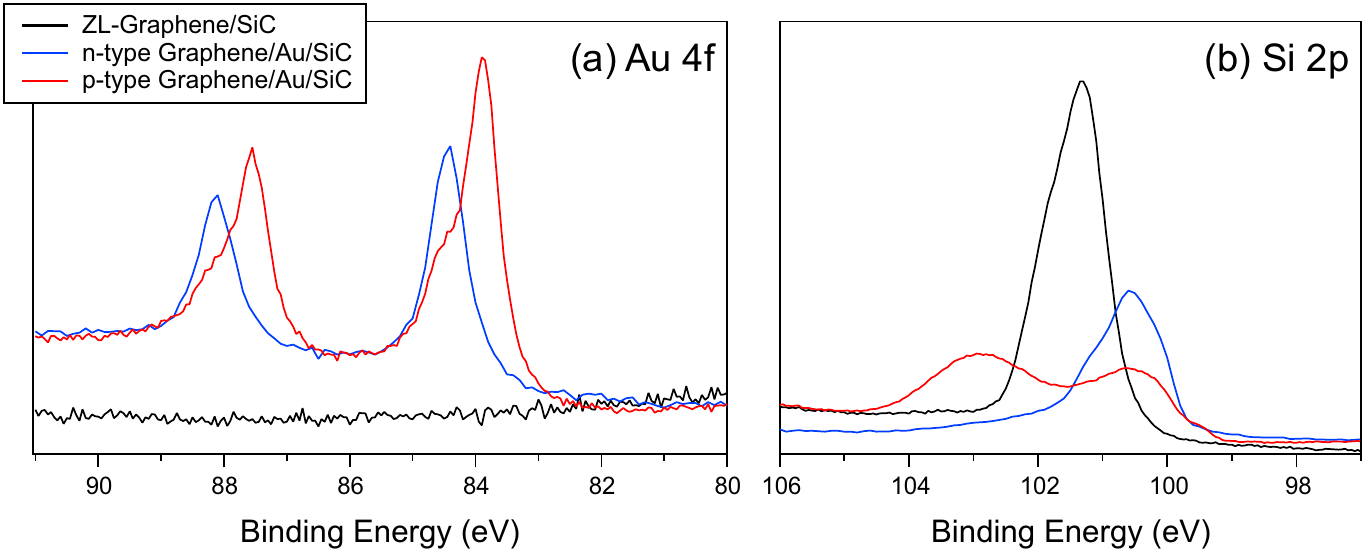} \caption{
Characterization of Au intercalation by (a) Au 4f and (b) Si 2p core-levels for zero-layer (ZL) graphene, n- and p-doped phases of graphene on gold. \hv = 130 eV.
} \end{figure}

\begin{figure}[h] \includegraphics[width=0.9\textwidth]{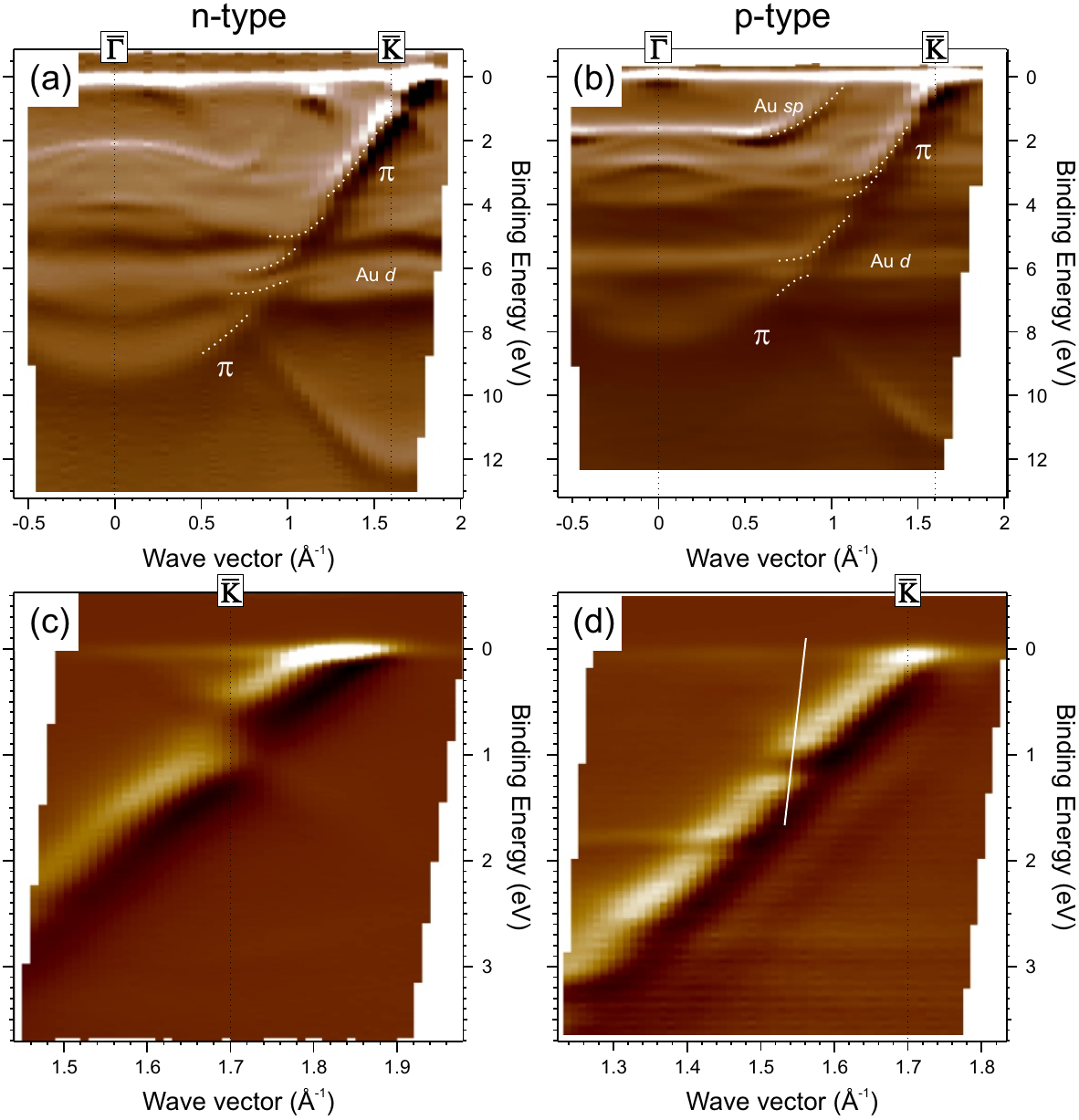} \caption{
Comparison of (a,c) n-type graphene/Au/SiC with low amount of Au and (b,d) p-type graphene/Au/SiC with high amount of Au. The hybridizations are very pronounced and are marked with white dashed lines. (c,d) The region near Fermi-level and \Kbar\ point shows (c) for n-type graphene a gapped Dirac point and (d) for p-type graphene various new hybridizations with flat bands. The white line on (d) shows the location in energy and momentum space of spin-resolved measurement presented on Fig. 3. On all (a-d) panels the first derivative of intensity over energy is presented. \hv = 62 eV. 
} \end{figure}

\begin{figure}[h] \includegraphics[width=0.5\textwidth]{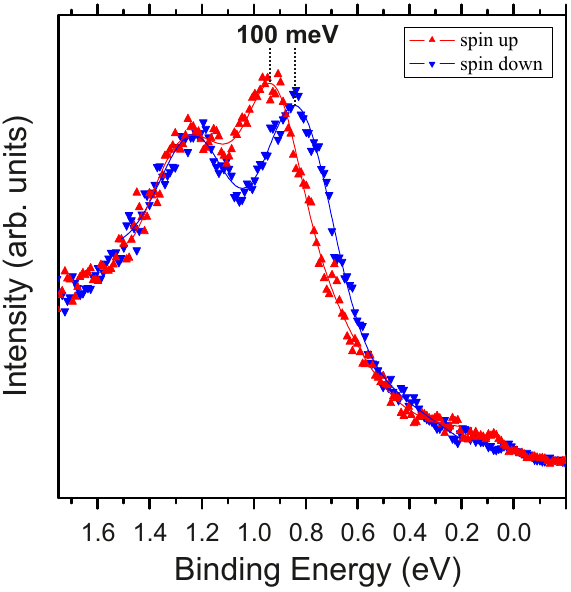} \caption{
Spin- and angle-resolved photoemission of p-type graphene near the hybridization region around 1 eV binding energy as indicated in Fig. 2d. The spin splitting is about 100 meV for the antibonding and undetectable for the bonding partner of the hybridization. \hv = 62 eV.
} \end{figure}

\end{document}